\documentclass[aps,prc,showpacs,twocolumn]{revtex4}
\usepackage{graphicx}
\usepackage{dcolumn} 
\usepackage{bm}
\usepackage{hyperref}
\usepackage{natbib}
\usepackage{amsmath}
\usepackage{multirow}
\begin{document}
\title{NMEs for $0\nu\beta\beta(0^+\rightarrow2^+)$ of two-nucleon mechanism for $^{76}$Ge}
\author{Dong-Liang Fang$^{a,b}$ and Amand Faessler$^{c}$} 
\address{{$^a$Institute of Modern Physics, Chinese Academy of Science, Lanzhou, 730000, China}}
\address{{$^b$University of Chinese Academy of Sciences, Beijing, 100049,China}}
\address{$^c$Institute for theoretical physics, Tuebingen University, D-72076, Germany}
\begin{abstract}
In this work we present the first beyond closure calculation for the neutrinoless double beta decay ($0\nu\beta\beta$) of $^{76}$Ge to the first $2^+$ states of $^{76}$Se. The isospin symmetry restored Quasi-particle random phase approximation (QRPA) method with the CD-Bonn realistic force is adopted for the nuclear structure calculations. We analyze the structure of the two nucleon mechanism nuclear matrix elements, and estimate the uncertainties from the nuclear many-body calculations. We find $g_{pp}$ plays an important role for the calculations and if quenching is included, suppression for the transition matrix element $M_{\lambda}$ is found. Our results for the transition matrix elements are about one order of magnitude larger than previous projected Hatree-Fock-Boglyubov results with the closure approximation.
\end{abstract}
\maketitle

\section{Introduction}
In the standard model, the nuclear weak decay is interpreted as the low-energy effective theory for weak interaction. This decay  is mediated by the left-handed gauge boson $W^{\pm}$. The mass of $W^{\pm}$ are acquired through the spontaneous symmetry breaking by the so-called Higgs mechanism. However, the Yukawa coupling of Higgs particle to neutrinos is absent in the standard model due to the absence of the right-handed neutrinos. The discovery of neutrino masses from oscillation experiments then asks for new physics beyond the standard model. As an extension to Standard model, the L-R symmetric model \cite{PS74,MP75,SM75} introduces the right-handed $SU(2)_R$ gauge symmetry and a hence heavy right-handed gauge boson from symmetry breaking with extra Higgs bosons at a higher energy scale beyond electroweak scale. In such a theory, the introduction of lepton number violating neutrino Majorana mass terms together with normal Dirac mass terms gives naturally the tiny neutrino mass through the so-called See-Saw mechanism \cite{Kin03}. Such extensions to the Standard Model could also affect the rare nuclear process called neutrinoless double beta decay ($0\nu\beta\beta$). The participation of right-handed weak gauge bosons will also induce the emission of right handed leptons. The simultaneous presence of weak currents with both chirality will introduce a momentum term into the neutrino propagator. These terms are not suppressed like the mass terms due to the smallness of neutrino mass. The right-handed weak currents will contribute to the decay to the ground states with extra terms and change the electron spectra \cite{SDS15}. Nevertheless, these terms are suppressed by the new physics parameters as well as the electron wave functions for p partial waves. Therefore, they are hindered in normal neutrinoless double beta decay compared to the neutrino mass mechanisms. On the other hand, the decay to the $2^+$ states, are dominated by the helicity changing mechanisms (V+A terms, see \cite{DKT85}). In this sense, the branching ratio of neutrinoless double beta decay to the $2^+$ state (hereafter $0\nu\beta\beta(2^+)$, the spin-parity of the final states of the decay are included inside the parenthesis ) could help to reveal the underlying mechanisms of this very rare decay. Nevertheless, experimentally such a process is extremely difficult to observe due to the large $2\nu\beta\beta(0^+)$ background around the position of $0\nu\beta\beta(2^+)$ Q value. Despite the difficulties, the observation of $0\nu\beta\beta(2^+)$ together with that for the decay to the ground states will determine the underlying mechanisms of the neutrinoless double beta decay and pave our way to new physics beyond the standard model. For example, the observation of $0\nu\beta\beta(2^+)$ could possibly rule out a category of mechanisms where no right-handed gauge bosons or fermions are present.

There are numerous publications dedicated to the nuclear many-body calculations for neutrinoless double beta decay with various approaches, {\it e. g.} the Shell Model calculations \cite{CNP99,Men17}, the QRPA calculations \cite{SPV99,SRF13}, the IBM calculations \cite{BKI13} and the nuclear meanfield calculations \cite{SYR14}, especially the recently developed {\it ab initio} methods \cite{YBE19}. However all these works focus on $0\nu\beta\beta(0^+)$, there aren't to many theoretical investigations available for $0\nu\beta\beta(2^+)$ in the literature for the matrix elements (NME). An earlier calculation \cite{Tom88} with the Projected Hartree Fock Bogoliubov method (PHFB) suggests, that the nuclear matrix elements (NME) for this decay mode are much smaller than for the decay to ground states. The QRPA method is widely used in double beta decay calculations \cite{SRF13,ME13,HS15}. The QRPA can also be adopted to describe the vibrational $2^+$ states. Therefore, there are attempts to use QRPA to calculate $\beta\beta$-decays to the first $2^+$ state \cite{SC93,SSF97}. And most of them focus on the $2\nu\beta\beta$ case. In this study, we go step forward by carrying out the QRPA calculations for $0\nu\beta\beta(2^+)$ for $^{76}$Ge. Taking advantage of the QRPA method, we take into account the contributions from all the intermediate states. Also we can include the isoscalar particle-particle residual interaction which is missing in PHFB calculations. At this first attempt, we do not include too many examples. We focus on one nucleus, $^{76}$Ge which is also the candidate treated in ref.  \cite{Tom88}. It has been shown in 
\cite{Tom88}, that besides the nucleon mechanism, the N* could also play an important role, we will not discuss this in the current study. Also, as  in \cite{CNP99}, the induced weak current will further reduce the NME, this will be neglected in this work. As well as the heavily suppressed neutrino mass mechanism for $0\nu\beta\beta(2^+)$ through nuclear recoil \cite{Tom99}.

The current article is arranged as follows: first we give the general formalism of our many-body calculations, and then we show  the results and discuss possible uncertainties, and finally we present the conclusions and outlook.

\section{Fromalisms}

The half-lives of $0\nu\beta\beta(2^+)$ can be expressed in a simple form,  while we consider the light neutrino only \cite{DKT85,Tom88}:
\begin{eqnarray}
\tau^{-1}=F_1(\langle \lambda \rangle M_\lambda - \langle \eta \rangle M_\eta)^2 + F_2 (\langle \eta \rangle M'_{\eta} )^2
\end{eqnarray}

Here $F_{1(2)}$ are the phase space factors expressed in \cite{DKT85,Tom88}. $\langle \lambda \rangle$ and $\langle \eta \rangle$ are the new physics parameters which are model dependent. In the L-R symmetric model \cite{MP75}: $SU(2)_L\times SU(2)_R\times U(1)_{B-L}$, we have \cite{DKT85}:
\begin{eqnarray}
\langle \lambda \rangle=\lambda\sum_{j} U_{ej} V_{ej} \quad \langle \eta \rangle =\eta \sum_{j} U_{ej} V_{ej}
\end{eqnarray}
Here $U_{ej}$ and $V_{ej}$ are matrix elements for the generalized PMNS matrix \cite{Xin11}. $\lambda\approx (M_{WL}/M_{WR})^2$ are the square of the ratio of the masses between the mass eigenvalues of the light and heavy gauge bosons, $\eta\approx \tan\xi$ is the mixing angle between the left-handed gauge boson and the heavy gauge boson mass eigenvalues.

 $M_{\lambda}$, $M_{\eta}$ and $M'_{\eta}$ are the nuclear matrix elments (NMEs) as combinations of different components \cite{DKT85}:
\begin{eqnarray}
M_{\lambda}=\sum_{i=1}^5 C_{\lambda i} M_i,  \quad 
M_{\eta}=\sum_{i=1}^5 C_{\eta i} M_i ,\quad
M'_{\eta}=\sum_{i=6}^7 C'_{\eta i} M_i
\end{eqnarray}
The different coefficients $C_{Ii}$ are given in table I of \cite{Tom88}.

The general form of above NMEs can be expressed as:
\begin{widetext}
\begin{eqnarray}
M_{I}&=& \sum_{J^\pi m_i m_f} \sum_{J' \mathcal{J} \mathcal{J}'}
\left\{
\begin{array}{ccc}
j_{p} & j_{p'} & \mathcal{J} \\
j_{n} & j_{n'} & \mathcal{J}' \\
J & J' & 2
\end{array}
\right\} \langle j_p j_{p'} \mathcal{J} || \mathcal{O}_I || j_n j_{n'} \mathcal{J}' \rangle
\nonumber \\
&\times& \frac{(-1)^{J'+J} }{\sqrt{2J'+1}} \langle 2^+_{1f} || \widetilde{[c_p^\dagger \tilde{c}_{n}]}_{J'} || J^\pi m_f \rangle \langle J^\pi m_f || J^\pi m_i \rangle \langle J^\pi m_i || [c_{p}^\dagger \tilde{c}_n]_J ||0^+_i \rangle
\end{eqnarray} 
\end{widetext}
A general derivation of the single particle matrix elements are given in the appendix. For the related operators, we have the form \cite{Tom88}:
\begin{eqnarray}
\mathcal{O}_1&=& \sigma_1\cdot\sigma_2[\hat{r}\otimes\hat{r}]^{(2)}h(r) \\
\mathcal{O}_2&=& [\sigma_1\otimes\sigma_2]^{(2)} h(r) \\
\mathcal{O}_3&=& [[\sigma_1\otimes\sigma_2]^{(2)}\otimes[\hat{r}\otimes\hat{r}]^{(2)}]^{(2)} h(r) \\
\mathcal{O}_4&=& [\hat{r}\otimes\hat{r}]^{(2)} h(r) \\
\mathcal{O}_5&=& [(\sigma_1+\sigma_2)\otimes[\hat{r}\otimes\hat{r}]^{(2)}]^{(2)} h(r)
\end{eqnarray}

And:
\begin{eqnarray}
\mathcal{O}_6&=& [(\sigma_1-\sigma_2)\otimes[\hat{r}\otimes\hat{r}]^{(1)}]^{(2)} \frac{r_+}{r}h(r) \\
\mathcal{O}_7&=& [(\sigma_1-\sigma_2)\otimes[\hat{r}\otimes\hat{r}]^{(2)}]^{(2)} \frac{r_+}{r}h(r)
\end{eqnarray}
Here $r$ is the relative distance between the two decaying nucleons $\vec{r}=\vec{r}_1-\vec{r}_2$ and $\vec{r}_+=(\vec{r}_1+\vec{r}_2)/2$ is their average position. 

Here, $M_1$, $M_2$ and $M_3$ are the space-space components of the current-current interaction, while $M_4$ is the time-time components and $M_5$, $M_6$ and $M_7$ are the time-space components. These time-space components of the NME's appear only in the L-R symmetric case and are missing in the neutrino mass mechanisms. In all these NME's, we find a similarity between $M_2$ and $M_{GT}^{0\nu}$ for $0\nu\beta\beta(0^+)$ as well as $M_3$ and $M_T^{0\nu}$. The GT operator $\sigma$ or the tensor operator $[\sigma\otimes \hat{r}]^2$ replace the scalar products in $0\nu\beta\beta(0^+)$. We also find an  analog similarity between $M_4$ and $M_F^{0\nu}$, where $\hat{r}$'s come from the p-wave electron and the momentum term form a tensor product instead of a scalar product in $0\nu\beta\beta(0^+)$.

The neutrino potential differs from that of the mass mechanism due to the momentum terms in the neutrino propagator \cite{SDS15}:
\begin{eqnarray}
h(r)=\frac{2 R}{\pi} r\int \frac{ F(q^2) j_1(qr) q^2 dq}{q+E_N}
\end{eqnarray}
By deriving this, we assume that the two electrons share the decay energy, therefore $E_N=E_m+M_m-(M_i+M_f)/2$ and $E_m$ is the excitation energy of the m$th$ excited states of the intermediate nucleus. The nuclear radius $R=1.2A^{1/3}\b [fm]$ introduced here makes the final NME dimensionless. For the form factor $F(q^2)$ we use a dipole form with the parameters are as in \cite{SPV99}.

Usually, an extra radial function $f(r)$ should be multiplied to the above expression to take into consideration the strong repulsive nature of nucleon-nucleon interaction at short range. This is usually called the short range correlation (src) function, and in our calculation we choose the CD-Bonn or Argonne src extracted from the corresponding nuclear force with the form given  in \cite{SFM09}.

For the reduced one-body density for transitions from intermediate states to final $2^+$, the expression is complicated \cite{FF20}:
\begin{widetext}
\begin{eqnarray}
\frac{\langle 2^+_f ||[\tilde{c}_p^\dagger c_{n}]_{J'} ||J^\pi m\rangle}{\sqrt{2J'+1}}&=&\sqrt{5(2J+1)}
[\sum_{ p'\le p}\frac{(-1)^{j_{p'}+j_n}}{\sqrt{1+\delta_{p p'}}}  
\left\{
\begin{array}{ccc}
2 & j_{p'} & j_p \\
j_{n} & J' & J
\end{array}
\right\}
(u_p u_n \mathcal{X}_{p' p}^{2^+_f} X^{J^\pi m}_{p'n}-v_p v_n \mathcal{Y}_{p' p}^{2^+_f} Y^{J^\pi m}_{p'n})
\nonumber \\ 
&+& \sum_{p'\ge p}\frac{(-1)^{j_p+j_n}}{\sqrt{1+\delta_{p p'}}} 
\left\{
\begin{array}{ccc}
2 & j_{p'} & j_p \\
j_{n} & 1 & 1
\end{array}
\right\}
(u_p u_n \mathcal{X}_{p p'}^{2^+_f} X^{m}_{p'n}-v_p v_n \mathcal{Y}_{p p'}^{2^+_f} Y^{m}_{p'n})
\nonumber \\  
&-&\sum_{n'\le n}\frac{ (-1)^{j_n+j_p}}{\sqrt{1+\delta_{nn'}}} 
\left\{
\begin{array}{ccc}
2 & j_{n'} & j_n \\
j_{p} & 1 & 1
\end{array}
\right\} 
(v_p v_n \mathcal{X}_{n'n}^{2^+_f} X^{m}_{pn'} -u_p u_n \mathcal{Y}_{n' n}^{2^+_f} Y^{m}_{p n'} ) 
\nonumber \\
&-&
\sum_{n' \ge n}\frac{(-1)^{j_{n'}+j_p}}{\sqrt{1+\delta_{n n'}}} 
\left\{
\begin{array}{ccc}
2 & j_{n'} & j_n \\
j_{p} & 1 & 1
\end{array}
\right\}
(v_p v_n \mathcal{X}_{n n'}^{2^+_f} X^{m}_{p n'} -u_p u_n \mathcal{Y}_{n n'}^{2^+_f} Y^{m}_{p n'} ) ]
\nonumber \\
\end{eqnarray}
\end{widetext}
Here $X$'s and $Y$'s are the amplitudes for pn-QRPA (proton-neutron Quasi-particle Random Phase Approximation) describing the intermediate states and $\mathcal{X}$'s and $\mathcal{Y}$'s are the amplitudes for CC-QRPA (Charge Conserving QRPA) describing the final $2^+$ state \cite{FF20}. And $u$'s and $v$'s are the BCS coefficients.

The reduced one body density for transitions from the initial states to the intermediate states can be expressed as \cite{SPV99}:
\begin{eqnarray}
\frac{\langle J^\pi m_i || [c_{p}^\dagger \tilde{c}_n]_J ||0^+_i\rangle}{\sqrt{2J+1}}=\sum_{pn}(u_p v_n X_{pn}^{J^\pi,m_i}+u_n v_p Y_{pn}^{J^\pi,m_i})
\end{eqnarray}
We also introduce the overlap between the initial and final intermediate states with the form \cite{SPV99}:
\begin{eqnarray}
\langle J^\pi m_f || J^\pi m_i \rangle=\sum_{pn}(X_{pn}^{J^\pi,m_i}X_{pn}^{J^\pi,m_f}-Y_{pn}^{J^\pi,m_i}Y_{pn}^{J^\pi,m_f})\nonumber\\
\times(u_p^{i}u_p^f+v_p^iv_p^f)(u_n^iu_n^f+v_n^iv_n^f)_f\langle BCS|BCS\rangle_i \nonumber \\
\end{eqnarray}
For simplicity, we set $_f\langle BCS|BCS\rangle_i\approx 1$.

The details of derivations of the BCS coefficients and respective QRPA amplitudes for the current work are presented in \cite{FF20}. 

\section{Results and discussion}

\begin{table*}[htp]
\begin{center}
\begin{tabular}{c|ccccc|cc|cc|c}
\hline
   					&	$M_1$	&	$M_2$	&	$M_3$	&	$M_4$	&	$M_5$	
					&$M_\Lambda$&	$M_\eta$
   		&	$M_6$	&	$M_7$	&	$M'_\eta$ \\
\hline
PHFB\cite{Tom88}			&	0.151	&	0.027	&	-0.002	&	-0.049	&	-0.004
					&	0.002	&	0.061	
		&	0.074	&	0.042	&	0.001		\\
 \hline
Baseline			&    0.705 		&   -0.253 		&   -0.046 		&   -0.153 		&   -0.048 
					&    0.150 		&    0.469
		&    0.527 		&   -1.270 		&    1.519 \\
\hline
$N_{max}=5$ &    0.629 &   -0.208 &   -0.014 &   -0.124 &   -0.069 
&    0.151 &    0.438
&    0.661 &   -1.369 &    1.688 \\
$N_{max}=7$ &    0.640 &   -0.256 &   -0.048 &   -0.145 &   -0.063 
&    0.121 &    0.439
&    0.643 &   -1.251 &    1.564 \\
\hline
w/o src &    0.701 &   -0.234 &   -0.049 &   -0.154 &   -0.051 
&    0.128 &    0.451
&    0.485 &   -1.182 &    1.410 \\
Argonne src &    0.705 &   -0.250 &   -0.046 &   -0.153 &   -0.048 
&    0.149 &    0.467
&    0.519 &   -1.261 &    1.505 \\
\hline
L.O. &    0.749 &   -0.347 &   -0.051 &   -0.154 &   -0.041 
&    0.228 &    0.540
&    0.823 &   -1.756 &    2.152 \\
w/o $F(q^2)$ &    0.695 &   -0.241 &   -0.047 &   -0.154 &   -0.050 
&    0.136 &    0.457
&    0.529 &   -1.272 &    1.521 \\
Closure Energy &    0.696 &   -0.267 &   -0.043 &   -0.144 &   -0.041 
&    0.177 &    0.472
&    0.522 &   -1.247 &    1.493 \\
\hline
$g_{pp}^{T=0}=0$ &    0.611 &   -0.169 &   -0.054 &   -0.161 &   -0.065 
&    0.029 &    0.376
&    0.540 &   -1.240 &    1.496 \\
$g_{pp}^{T=1}=0$ &    0.795 &   -0.246 &   -0.034 &   -0.156 &   -0.034 
&    0.206 &    0.516
&    0.501 &   -1.437 &    1.665 \\
\hline
$g_A=0.75 $ &    0.695 &   -0.241 &   -0.047 &   -0.154 &   -0.050 
&    0.008 &    0.317
&    0.529 &   -1.272 &    1.249 \\
 \hline
\end{tabular}
\end{center}
\caption{The NME values for $0\nu\beta\beta(2+)$. Here the baseline calculation is explained in text. And also various approximations and parameters will be discussed in text.}
\label{rpar}
\end{table*}

\begin{figure*}
\begin{center}
\includegraphics[scale=0.45]{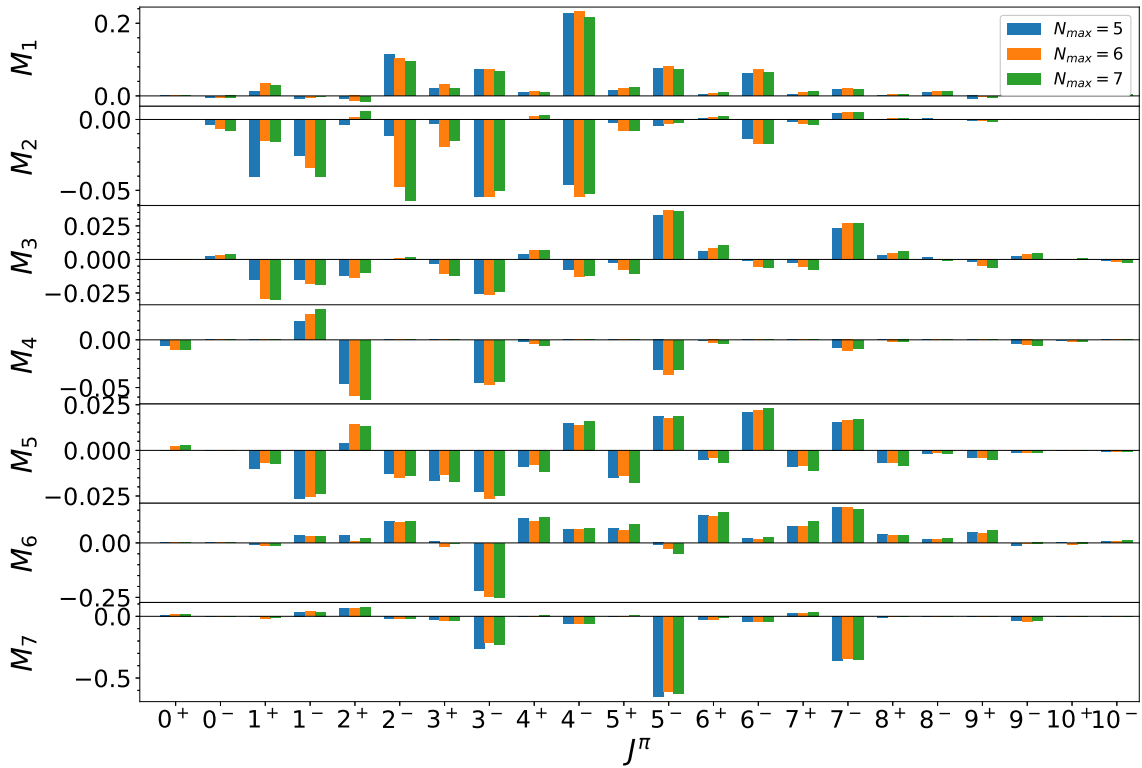}
\caption{(Color online) The dependence of the NMEs on the model space for different multipoles. Here $N_{max}$ refers to the largest principle quantum number for the outermost shell.}
\label{space}
\end{center} 
\end{figure*}

\begin{figure*}
\includegraphics[scale=0.45]{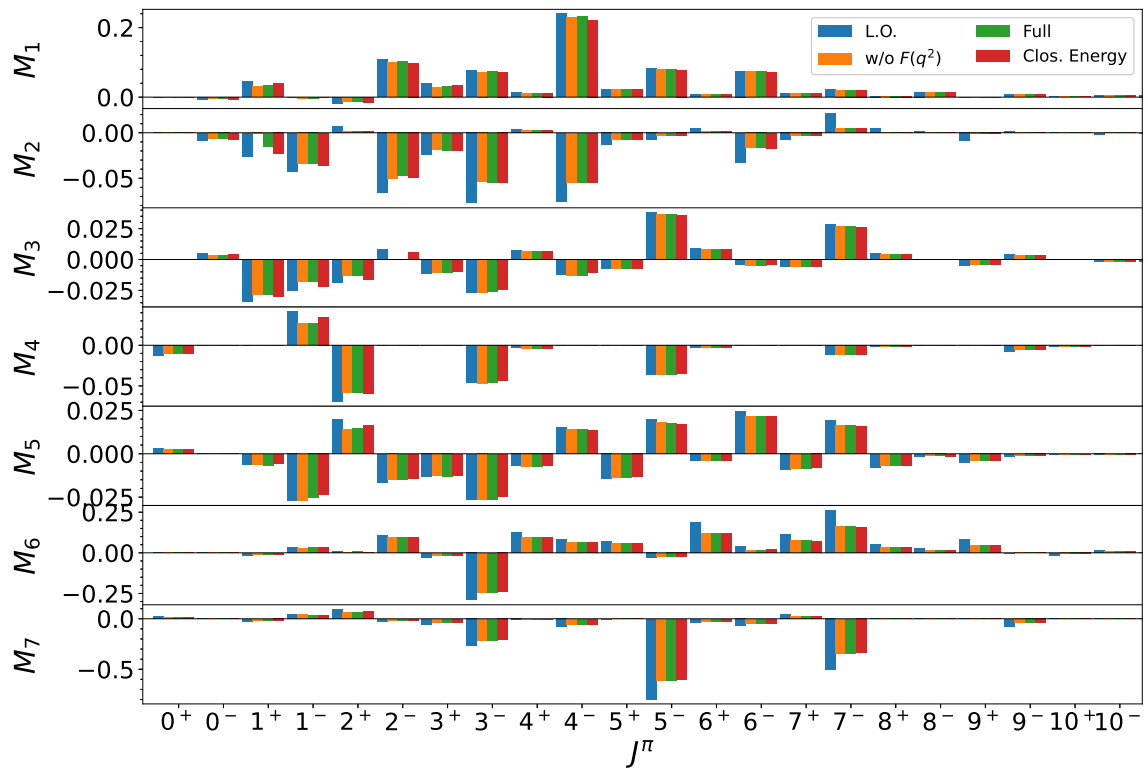}
\caption{(Color online) The NMEs for a Coulomb type neutrino potential (blue bars). The orange bars are NMEs without form factors and red bars are with excitation energies replaced by a closure energy. The green bar are our baseline calculations explained in the text.}
\label{corr}
\end{figure*}

\begin{figure*}
\includegraphics[scale=0.45]{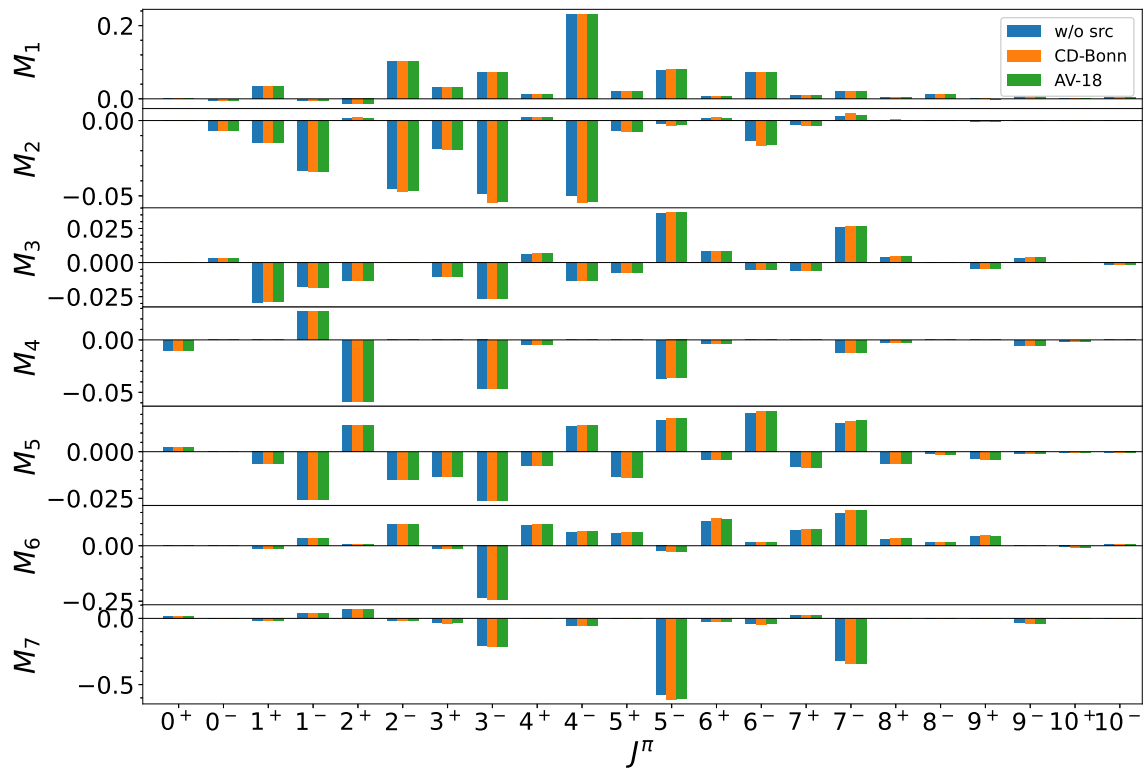}
\caption{(Color online) The NME dependence on the short range correlations.}
\label{src}
\end{figure*}

\begin{figure*}
\includegraphics[scale=0.45]{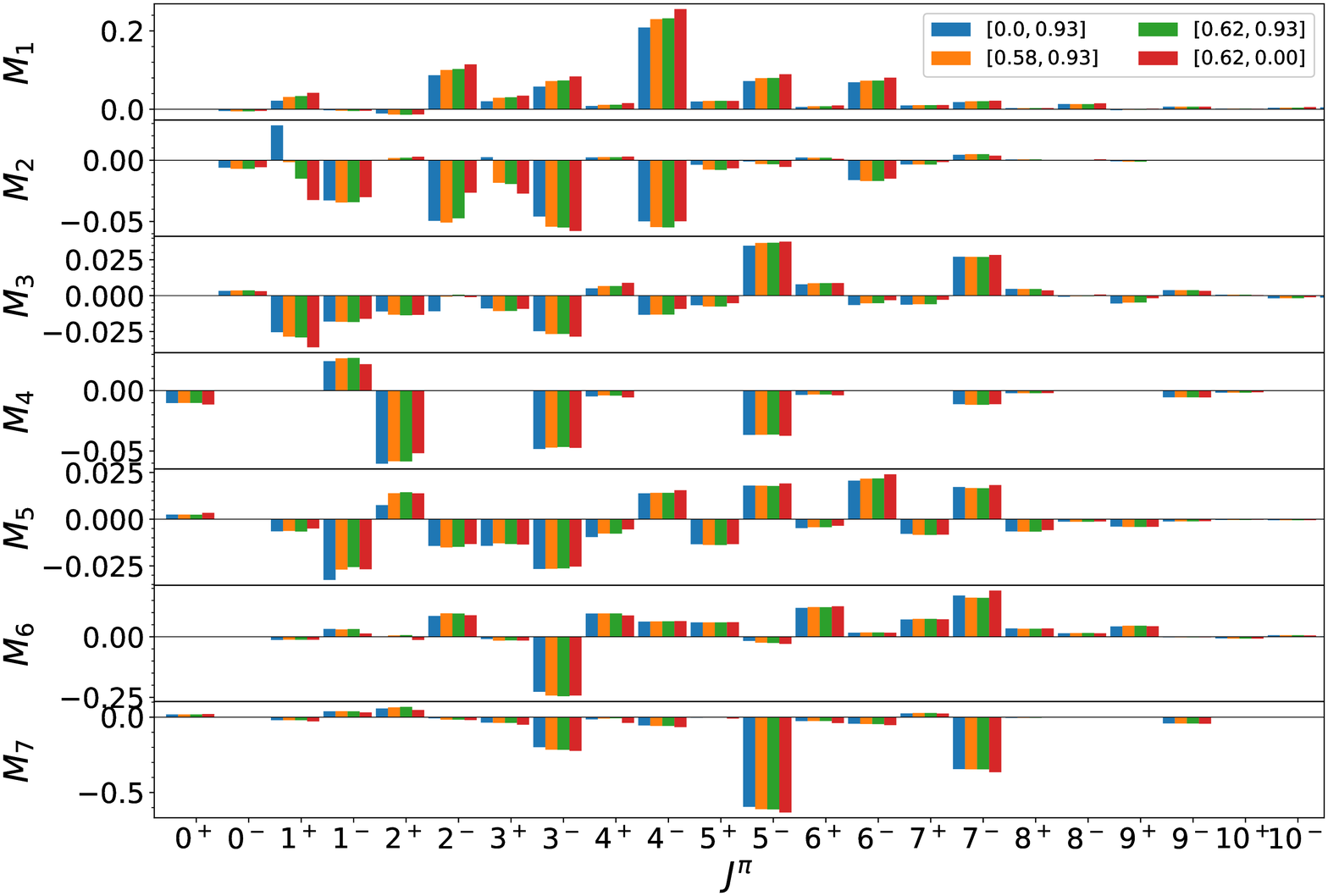}
\caption{(Color online) The NME dependence on $g_{pp}$'s. The values in the bracket are $g_{pp}^{T=0}$ and $g_{pp}^{T=1}$ respectively. The blue bars are results with $g_{pp}^{T=0}=0$, and the orange bars are $g_{pp}^{T=0}$ values which reproduce the $2\nu\beta\beta$ NME with $g_A=0.75g_{A0}$. The red bars are results with $g_{pp}^{T=1}=0$. And the green bars here again are the baseline calculations.}
\label{pp}
\end{figure*}

For our QRPA calculations, the single particle energies are taken from the solutions of Schr\"odinger equations with a Coulomb corrected Woods-Saxon potential. For the single particle wave functions, we use the Harmonic Oscillator wave functions. For the pairing part, we use the realistic CD-Bonn force derived from Br\"uckner G-matrix. This is also used for the pn-QRPA and CC-QRPA residual  interactions. A fine tuning of the interactions is needed to reproduce the experimental values. For the pairing part, we fit the two parameters $g_{pair}^p$ and $g_{pair}^n$ to reproduce the odd-even mass staggering. For pn-QRPA, we multiply the G-matrix by overall renormalization factors $g_{ph}$ and $g_{pp}$'s for particle-hole and particle-particle parts, respectively. We set $g_{ph}=1$. And for $g_{pp}$, we fit the iso-scalar channel ($g_{pp}^{T=0}$) and iso-vector channel ($g_{pp}^{T=1}$) separately. $g_{pp}^{T=0}$ is fixed by reproducing the experimental $2\nu\beta\beta$ and $g_{pp}^{T=1}$ are fixed to put to zero  the $2\nu\beta\beta$ Fermi matrix elements due to isospin symmetry restoration \cite{RF11}. For more details, one can refer to the our previous work \cite{FF20}. For a baseline calculation, we consider the CD-Bonn src  bare axial vector coupling constant $g_{A0}=1.27$. And we use a model space with $N_{max}=6$ which consists of 28 single particle levels for both neutrons and protons.

In table.\ref{rpar}, we present the NMEs for each operator. As a comparison, we also present the results from PHFB calculations. The current results (the baseline results) differ from the PHFB results by factors from five to more than one order of magnitude case by case. The largest deviation we see in  $M_7$ for the $M'_\eta$ part. We obtain  $M_7$ much larger than $M_6$ in magnitude.  Our results have have also different phases for these two NMEs, this then combined with the $C'_\eta$ coefficients  leads to the enhancement instead of the cancellations of $M'_{\eta}$. Therefore, our $M'_\eta$ is much larger than a previous PHFB calculation. PHFB  gets an approximate cancellation between $M_6$ and $M_7$, which leads to an negligible $M'_\eta$. 

For the $M_{\lambda(\eta)}$ part, we find the NME's have basically similar phases as the PHFB results except for $M_2$. On the other hand, our results are about one order of magnitude larger, although the relative ratios among different NMEs ($M_1-M_5$) are similar. Of these NMEs, $M_1$ is the largest. The second largest is $M_2$ and third is $M_4$. $M_3$ and $M_5$ are relatively small and hence less important. For $M_\lambda$, if we multiply the NMEs with the corresponding $C_{\lambda}$'s, we find that $M_1$ and $M_2$ contributes coherently, they are then cancelled by $M_4$, while the rest two NME's contributes less than 10\%. For $M_{\eta}$, all these three NMEs gives additive contributions, this makes $M_{\eta}$ about three times larger than $M_{\lambda}$. This is the reason, why  $M_{\eta}$ is larger than $M_\lambda$ as also observed in \cite{Tom88}. In our case, this ratio is about three. This is similar to PHFB calculations, however, in their calculation, the strong cancellation gives a negligible $M_{\lambda}$, the ratio is about 30. In short, in their calculation, only $M_{\eta}$ is important while $M_{\lambda}$ and $M'_{\eta}$ can be neglected due to the cancellations between different parts of the NME. We get quite different results, and $M'_{\eta}$ is the most important contribution, with a value about three times larger than $M_{\eta}$. This will affect the constraints on new physics models and needs further investigation.  

As we show above, no obvious suppression of $0\nu\beta\beta(2^+)$ NME's as claimed by \cite{Tom88} is found from current calculations compared to $0\nu\beta\beta(0^+)$, this also agrees with the $q$ terms in $0\nu\beta\beta(0^+)$ calculations \cite{MBK89}. This is the major difference of the  current work and \cite{Tom88}. This is also different from $2\nu\beta\beta(2^+)$, where the NME is suppressed by the cubic dependence of the energy denominator \cite{DKT85}. In this sense, suppression of $0\nu\beta\beta(2^+)$, if it exists, must be related to other issues. This may come from the uncertainties of the many-body approaches, such as the size of the model space or other structure ingredients for the  $2^+$ states which may lead to different transition rates from the intermediate states for various transitions. A more thorough comparative study could give us mor detailed  hints. 

Compared to previous calculations with PHFB \cite{Tom88}, the QRPA calculation goes beyond the closure approximation. We calculate explicitly the contribution from each intermediate state. In fig. (\ref{space}-\ref{pp}), we present the individual contributions from different multipoles of the intermediate states and we will also show how the different approximations may affect these results. In each graph, the results are compared with our present baseline calculations with standard conditions described above.

For details of the structure of the nuclear MME's, we start with our baseline calculation ({\it e.g.} orange bars in fig.\ref{space}). For $M_1$, the multipoles give positive contributions with several exceptions. Unlike $0\nu\beta\beta(0^+)$ ($M_{GT}^{0\nu}$), where the largest contributions come from low-spin intermediate states and the NME values decrease as spins increase, $M_1$ has its largest contribution from $4^-$. We find a rough trend that the NME's first increase and then decrease as spin increases. And as one would expect, the NME's from states with very high spin can be safely neglected. $M_2$ has basically the same characters as $M_1$ except the much smaller magnitude. A large contribution from $1^-$ is observed for $M_2$ but not for $M_1$.  For most multipoles, $M_2$ has different signs as $M_1$, this contradicts conclusions in \cite{Tom88}. Not all multipoles contribute equally with similar spins, we find that the states with negative parity generally contribute more. In some sense, these two NME's behave like $M^{0\nu}_{GT}$ for $0\nu\beta\beta(0^+)$ as we mentioned above. 
The smallness of final $M_3$ comes partially from its magnitude and partially from the cancellations between low-spin and high-spin multipoles. This is analog to $M^{0\nu}_T$, although these two NME's depend differently on $\hat{r}$.

$M_4$, the time-time component of the NME is on the other hand, very close to $M_{F}^{0\nu}$. The two NME's have one thing in common: Only intermediate states with natural parity ($\pi=(-1)^J$) have non-zero contributions. In the current calculation, all multipoles contribute negatively except $1^-$. All these multipoles contribute basically with the same magnitude as $M_2$ and $M_4$ is the major cancellation to $M_\lambda$. 

For the space-time components, $M_5$, $M_6$ and $M_7$, we find a quite different behavior. $M_5$ from each multipole is about one order of magnitude smaller than $M_6$ and $M_7$. Due to the strong cancellations among different multipoles,  $M_5$ is one of the smallest of all the NME's. Almost all multipoles contribute positively except a strong cancellation from $3^-$. For $M_7$, the important contributions comes from three multipoles ($3^-$, $5^-$ and $7^-$), all the other multipoles contribute much less. The  lack of cancellations from other multipoles thus makes $M_7$ the largest from all the NME's.

A possible cause of the smallness of NMEs in ref. \cite{Tom88} may come from the small model space used whith only  two major shells. To test this, we plot in fig.\ref{space} the results with $N_{max}=5$ (blue bars) and $N_{max}=7$ (green bars). The sensitivity of the NME's to diffeent  model spaces are different. Also different is the sensitivity of the individual multipoles for each NME. For $M_1$, the influence from model space is generally small, there is no unique trend for different multipoles under the change of the model space. For some multipoles, the NME decreases with enlarged model space, but most cases we find that the extra orbitals will first enhance but then reduce the NME. The orbitals of different parity contribute to the NME differently. The addition of N=6 shell brings in the positive parity orbitals which enhance the NMEs for multipoles such as $3^-$ or $5^+$, while the negative parity orbitals from N=7 shells reduce the NME's. Unlike the case of $M_1$, the increase of the model space enlarge $M_2$, especially for $1^+$, $2^-$ {\it etc}. For $1^+$, a strong suppression is observed when the N=6 shell is added. But such addition gives strong enhancement to $2^-$ or $3^+$. The addition of the N=7 shell to the model space causes milder changes. We see enhancement from $1^-$ and $2^-$ but reductions from $3^-$ and $4^-$ intermediate states. A similar behavior shows the  $M_3$, where we find a strong enhancement from $1^+$ too. For all other multipoles, the change due to model space enlargement is relatively small. For $M_4$, low spin multipoles are much more sensitive to the model space changes and different multipoles behave differently, though we cannot find any specific patterns. This is also true for the space-time components of the NME. In general, they are less affected by the change of model space in our calculations. 

If we look at the total changes of each NME from $N_{max}=5$ to $N_{max}=6$, $M_1$ increases about $10\%$. this is the smallest change among all the NMEs. Meanwhile, all the other NMEs changes about $20\%$ or more. By percentage $M_3$ changes by more than $200\%$. By the absolute values of NMEs, $M_1$, $M_2$, $M_3$ and $M_4$ get enhanced while the rest get reduced. When the $N=7$ shell is added, the change is relatively milder, especially for $M_2$, $M_3$ and $M_4$, this implies a general trend of convergence of the results with a larger model space. But for $M_1$, $M_5$, $M_6$ and $M_7$, we find a slower trend of convergence than for the above NMEs. In either cases, the changes from adding the $N=7$ shell is smaller than adding the $N=6$ shell. Current results suggest that the errors of adopting the current model space are generally smaller than $20\%$. Also, these results suggest that the deviation of our calculations and those in \cite{Tom88} is not caused by the small model space, which  they adopted. It is most probable that the smallness of their results are caused by different nuclear structures. Additional  work is needed to clarify this.  

 
The form factor with a dipole form is widely used in $0\nu\beta\beta(0^+)$ calculations \cite{SPV99}. In our calculations, we use the same form for $g_V(q^2)$ and $g_A(q^2)$. We find that the form factors are not very important except for the $1^+$ states of $M_2$. This may suggest that with the actual neutrino potential the low momentum parts where the $q^2$ satisfies $g_A(q^2)\sim g_{A}(0)$ dominate, and the high momentum parts are either small or cancels each other. A careful check suggests the latter should apply. These behaviors also helps to explain the large reduction for some NME's when the realistic neutrino potential is considered. With low momenta, the intermediate energies become important. In this sense, the choice of the intermediate energies  becomes important. For calculations with closure approximation, a closure energy is needed. So we also check how the use  of closure energy would affect the NME's. This is illustrated with the red bars in fig.\ref{corr}. A  closure energy of $7$ MeV is used. We find for most multipoles of most NME's, the proper choice of closure energy brings small errors to the calculation and we can draw the conclusion, that using the closure energy in $0\nu\beta\beta(2^+)$ barely changes the final results, the errors are within several percents, this agrees with $0\nu\beta\beta(0^+)$ QRPA calculations. Another important correction comes from the induced weak current \cite{SPV99}, it is not included in the current calculations and will be implemented in our future study. If that is included, the NME will most probably be further reduced and new potentials needs to be introduced \cite{SDS15}.

In $0\nu\beta\beta(0^+)$, src only plays an important role for the heavy neutrino mass mechanism \cite{FFS18,SYR14}, while for the light neutrino mass mechanism, the correction is relatively small, up to only several percents \cite{FFS18,SYR14}. In current calculations, we adopt two src's \cite{SFM09}: the CD-Bonn type and Argonne type, they are obtained by fitting the respective nuclear potentials. From fig.\ref{src}, we find similar trends as $0\nu\beta\beta(0^+)$, the general correction from src is about several percent and for most cases, the two src's give similar amount of corrections. For almost all multipoles, we find that the src enhances the NMEs more or less. However, the net effects to the NMEs are slight reductions for $M_3$ and $M_5$ due to cancellations among different multipoles. And enhancements of NME values for other operators are observed. The largest correction comes from $M_6$, then $M_7$ and $M_2$. For $M_6$ we can also find slight difference between the two src's. 

The most important parameters in QRPA calculations are the particle-particle interaction strength $g_{pp}$'s. For $0\nu\beta\beta(0^+)$, one finds that $M_{GT}$ for both $2\nu\beta\beta$ and $0\nu\beta\beta$ depends sensitively on the isoscalar strength $g_{pp}^{T=0}$ while $M_{F}^{0\nu}$ are sensitive to the isovector strength $g_{pp}^{T=1}$. So we also test such dependence in current calculations. As we have shown in \cite{FF20} for the case of $2\nu\beta\beta$ to $2^+$ states, the GT type decays are also sensitive to $g_{pp}^{T=1}$ since the isovector particle-particle residue force affects the structure of $2^+$ states in QRPA calculations. In fig.\ref{pp}, we first switch off the isoscalar interaction (the blue bar). Then the effects of this interaction to NMEs can be estimated by comparing the blue and green bars. The results suggest that, while $M_6$ and $M_7$ are not closely related to $g_{pp}^{T=0}$, $M_2$ comes out to be the most sensitive one like its $0\nu\beta\beta(0^+)$ counterpart $M_{GT}^{0\nu}$. Similar as $M_{GT}^{0\nu}$, $1^+$ comes out to be the most sensitive multipole, the increase of the $g_{pp}^{T=0}$ drastically changes the NME most probably due to the SU(4) symmetry restoration\cite{RF11}. And the effects of isoscalar residue interactions to the specific multipoles of specific NMEs are quite different, some NMEs are enhanced and some are reduced. In total, the introduction of isoscalar interactions enhances $M_1$, $M_2$ and $M_7$, but reduces other NMEs. And the magnitudes of these enhancements and reductions are really case dependent. The similar thing happens to isovector interactions (this interaction has been switched off for the red bars in fig.\ref{pp}), for some cases they are more important than the isoscalar interactions. In general, the particle-particle interaction is one of the most important source of the errors for QRPA calculations. $M_4$ is at least sensitive to $g_{pp}$'s while $M_2$ and $M_7$ are really sensitive to $g_{pp}^{T=0}$ and $g_{pp}^{T=1}$ respectively. Especially $M_2$, the absence of isoscalar interactions will reduce the NME by more than $30\%$. 

The special attention should be paid to the case of the quenched $g_A$. In all our above analysis, we assume that $g_A$ is not quenched, however, in nuclear medium quenching of $g_A$ is observed. In current calculations, we use the simplest treatment for quenching, that is simply change the value of $g_A$, while there exists more fundamental approaches for the quenching of $g_A$  using the chiral two-body currents\cite{ESV14}. In our case, the difference between quenched $g_A$ calculations and our baseline calculations are the different fit of $g_{pp}^{T=0}$. Therefore, the difference for the individual NMEs are small as in fig.\ref{pp}. But the quenched $g_A$ will also change the coefficients $C$'s. As a result, the NMEs $M_{\lambda}$, $M_{\eta}$ and $M'_{\eta}$ have been changed too. In general, these NME's are reduced due to the convention used in\cite{Tom88}. For $M'_{\eta}$ a reduction of a rough factor of $g_A/g_{A0}$ is expected, since the two components $M_6$ and $M_7$ has the same dependence on $g_A$. For $M_\lambda$ and $M_\eta$, different components have a different $g_A$ dependence. And if we take $g_A=0.75g_{A0}$ we find a rough cancellation for $M_\lambda$ as predicted in \cite{Tom88}, this comes from the interplay among different components. Meanwhile the reduction for $M_\eta$ is basically $25\%$. The severe reductions of $M_\lambda$ emphasize the importance of where the quenching of $g_A$ originates and how to treat it properly.

\section{Conclusion and Outlook}
In this study, we calculate the NME of $0\nu\beta\beta$ to $2^+_1$ for $^{76}$Ge. We got quite large results compared to previous calculations. We estimate the errors of current calculation by changing several parameters we use. We find that $g_A$ may be a very important issue for the final NME's. Further investigations, such as the role of the induced weak current, the anharmonicity beyond QRPA, and the decay mechanism mediated by N*, are needed for much more detailed conclusions.

\section*{acknowledgement}
This work is supported by the "Light of West China" program and the "From 0 to 1 innovative research" program both from CAS.  

\begin{appendix}
\section{Derivation of single particle matrix elements in particle-particle channel}
The seven decay operators are taken from \cite{Tom88} and presented above. They can be written in the forms of combination of three parts: the relative coordinate ($\mathcal{O}^r(\vec{r})\equiv \mathcal{O}_{J}(\hat{r}) h(r)$), the center of mass coordinate ($\mathcal{O}^R(\vec{R})\equiv \mathcal{O}_J(\hat{R}) f(R)$) and spin part ($\mathcal{O}^{S}(\vec{\sigma}_1,\vec{\sigma}_2)$). These operators can then be expressed in a general form $\mathcal{O}^{(2)}_I=[[O_{J_1} (\hat{r}) \otimes O_{J_2} (\hat{R})]^{(J')}\otimes O_{J_3}(\vec{\sigma}_1,\vec{\sigma}_2)]^{(2)}h(r)f(R)$.

We assume the quantum numbers for the single orbital are $(n_1,l_1,j_1)$ and $(n_2,l_2,j_2)$ for protons and $(n_1',l_1',j_1')$ and $(n_2',l_2,j_2')$ for neutrons. The single matrix elements of these operators under harmonic oscillator basis can be expressed as:
\begin{eqnarray}
&&\langle p p' \mathcal{J} || [[O_{J_1} (\hat{r}) \otimes O_{J_2} (\hat{R})]^{(J')}\otimes O_{J_3}(\vec{\sigma}_1,\vec{\sigma}_2)]^{(2)} ||n n' \mathcal{J}'\rangle \nonumber \\
&=&\sum_{nl\mathcal{N}\mathcal{L}L S} A_{pp'\mathcal{J},LS}  \langle n_{1} l_1, n_2 l_2,L | nl \mathcal{N}\mathcal{L},L\rangle \nonumber \\
&\times&\sum_{n'l'\mathcal{N}'\mathcal{L}'L' S'}A_{nn'\mathcal{J}',L'S'}\langle n_{1}' l_1', n_2' l_2',L' | n'l' \mathcal{N}'\mathcal{L}',L'\rangle \nonumber\\
&\times&\langle n l \mathcal{N}\mathcal{L} L; s_1,s_2,S; \mathcal{J}|| \mathcal{O}_I^{(2)} || n' l' \mathcal{N}'\mathcal{L}' L'; s_1',s_2',S'; \mathcal{J}'\rangle
\end{eqnarray}
Here $A_{pp'(nn')J,LS}$ is the 9j-symbol for JJ to LS coupling transformation:
\begin{eqnarray}
A_{\tau\tau'J,LS}&=&(2S+1)(2L+1)\sqrt{(2j_\tau+1)(2j_{\tau'}+1)} \nonumber \\
&\times&\left\{
\begin{array}{ccc}
\frac{1}{2} & l_\tau & j_{\tau} \\
\frac{1}{2} & l_{\tau'} & j_{\tau'} \\
S & L & J
\end{array}
\right\}
\end{eqnarray}

And $\langle n_{1} l_1, n_2 l_2,L | nl \mathcal{N}\mathcal{L},L\rangle$ is the Brody-Moshinski transformation coefficients \cite{Mos59}. 
 
Using techniques from {\it e.g.} \cite{Edm96}, we could further get the expressions of each operator:
\begin{eqnarray}
&&\langle n l \mathcal{N}\mathcal{L} L; s_1,s_2,S; \mathcal{J} || \mathcal{O}_I^{(2)} || n' l' \mathcal{N}'\mathcal{L}' L'; s_1',s_2',S'; \mathcal{J}'\rangle \nonumber \\
&=&\sqrt{5(2\mathcal{J}+1)(2\mathcal{J}'+1)}
\left\{
\begin{array}{ccc}
L & L' & J' \\
S & S' & J_3 \\
\mathcal{J} & \mathcal{J}' & 2
\end{array}
\right \}
\nonumber \\
&\times& \sqrt{(2\mathcal{J}+1)(2\mathcal{J}'+1)(2J'+1)}
\left\{
\begin{array}{ccc}
l & l' & J_1 \\
\mathcal{L} & \mathcal{L}' & J_2 \\
L & L' & J'
\end{array}
\right \}
\nonumber \\
&\times&\langle nl ||O_{J_1} (\hat{r}) f(r)||n'l'\rangle \langle \mathcal{N}\mathcal{L} || O_{J_2} (\hat{R}) f(R) ||\mathcal{N}'\mathcal{L}'\rangle 
\nonumber \\
&\times& \langle s_1,s_2,S ||O_{J_3}(\vec{\sigma}_1,\vec{\sigma}_2) ||s_1',s_2',S'\rangle
\end{eqnarray}
All these reduced matrix elements with different $O_{J_1}(\hat{r})$ {\it etc.} can be calculated analytically in harmonic oscillator basis and we omit their derivations in current article, one could refer to references such as \cite{Edm96}.

\end{appendix}

\end{document}